\documentclass[a4paper]{jpconf}
\usepackage{graphicx,amsmath,amssymb,bm}
\newcommand{\kf}{k_{\rm F}}
\newcommand{\fmiq}{\, \text{fm}^{-3}}
\newcommand{\mev}{\, \text{MeV}}

\newcommand{\znbbeq}{0\nu\beta\beta}
\newcommand{\znbb}{$\znbbeq$}

\begin{document}
\title{Low-energy Electro-weak Reactions}

\author{Doron Gazit}

\address{Racah Institute of Physics and
The Hebrew University Center for Nanoscience and Nanotechnology, 
The Hebrew University, 91904 Jerusalem, Israel}

\ead{doron.gazit@mail.huji.ac.il}

\begin{abstract}
Chiral effective field theory (EFT) provides a systematic and controlled approach to low-energy nuclear physics.
Here, we use chiral EFT to calculate low-energy weak Gamow-Teller transitions. We put special emphasis on the role of two-body (2b) weak currents within the nucleus, and discuss their applications in predicting physical observables.
\end{abstract}

\section{Introduction}
Electro-weak nuclear reactions at low energies, which are the subject of this contribution, have an important role 
in many physical scenarios.  They are the microscopic motor of astrophysical phenomena, such as 
solar fusion, core collapse 
supernovae and the nucleosynthesis within. In addition, these reactions are used to study nuclear structure and 
dynamics. Moreover, low-energy $\beta$-decays and weak transitions provide a window into high-energy physics. 
For example, superallowed decays allow high precision tests of the Standard Model,
and \znbb\ decays probe the nature of neutrinos, their hierarchy and
mass. High accuracy theoretical, as well as experimental understanding of the nuclear response is a key ingredient needed to accomplish these tasks. 

\begin{figure}
\begin{center}
\includegraphics[scale=.7,clip=]{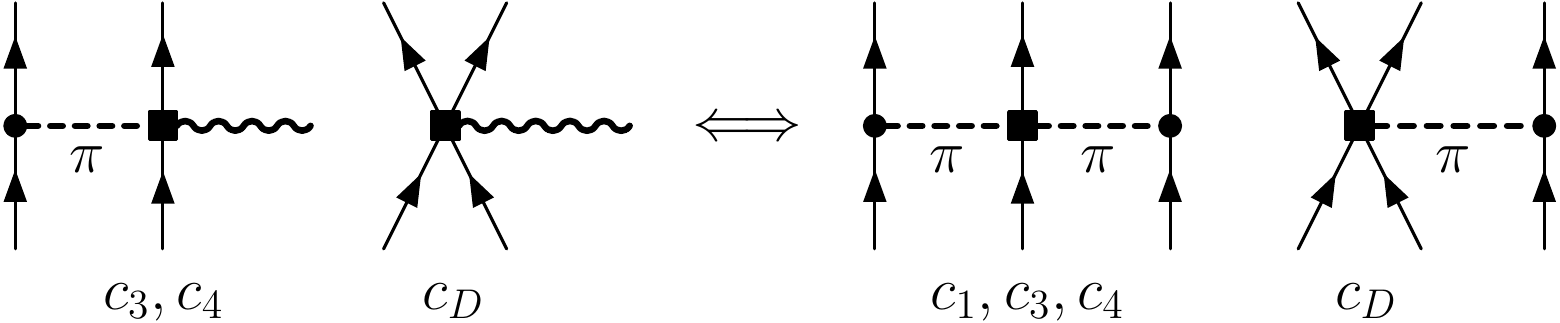}
\vspace*{-4mm}
\caption{In chiral EFT, Feynman diagrams of chiral 2b currents (left) correspond to 3N force contributions (right) that depend on the same low energy constants.}
\label{currents}
\vspace*{-3mm}
\end{center}
\end{figure}

However, the non-perturbative character of QCD at low energies makes a theoretical calculation of the nuclear response extremely challenging. In the last two decades, chiral effective field theory of QCD at low energies was developed, enabling a construction
of a nuclear Lagrangian that is consistent with QCD. Chiral EFT provides a systematic basis for nuclear forces and
consistent electroweak currents~\cite{RMP,Park}, where pion couplings
contribute both to the electroweak axial current
and to nuclear interactions. This is already seen at leading
order: the axial constant, $g_A$, determines the axial one-body (1b) current and the
one-pion-exchange potential. Two-body (2b) currents, also known as
meson-exchange currents, enter at higher order, just like
three-nucleon (3N) forces~\cite{RMP}. As shown in Fig.~\ref{currents},
the leading axial contributions are due to long-range one-pion-exchange and
short-range parts~\cite{Park}, with couplings $c_3,\, c_4$ and $c_D$,
which also enter the leading 3N (and subleading NN) 
forces~\cite{RMP,Gazit}. 

In this contribution we present some recent studies of chiral currents and their consistency with nuclear forces. 
We emphasize the importance of using chiral 2b currents in weak Gamow-Teller transitions in nuclei.

\section{Gamow-Teller transitions}
At low energies, the coupling of weak probes to a nucleus is given by
the current-current interaction, $H_W = \frac{G_F}{\sqrt{2}} \int d^3{\bf r}
\, e^{-i {\bf p} \cdot {\bf r}} j_{L \mu} J_L^{\mu \dagger} + {\rm
h.c.}$, where $G_F$ is the Fermi constant, ${\bf p}$ the momentum
transferred from nucleons to leptons, and $j_{L \mu}$ the leptonic
current of an electron coupled to a left-handed electron
neutrino. This approximation creates a separation between the leptons and baryons,
which is a great simplification of the problem. It suggests that only the symmetry of the interaction 
is needed to determine the structure of the current. For example, any axial probe, 
regardless of its fundamental origin,
would couple to the same current within the nucleus, which is the N{\"{o}}ther current derived 
from the axial symmetry of the chiral Lagrangian, which we will consider up to next-to-next-to-next-to 
leading order (N$^3$LO)~\cite{Park}.

The nuclear current $J_L^{\mu
\dagger}$ is organized in an expansion in powers of momentum $Q\sim
m_\pi$ over a breakdown scale $\Lambda_{\rm b} \sim 500
\, \rm{MeV}$. To order $Q^2$ (and also $Q^3$ in this counting), the 1b current,
$J_L^{\mu\dagger}({\bf r}) = \sum_{i=1}^{A} \tau_{i}^{-} \bigl[
\delta^{\mu0} J_{i,{\rm 1b}}^{0} - \delta^{\mu k}J_{i,{\rm 1b}}^{k}
\bigr] \delta({\bf r}-{\bf r}_i)$,
has temporal and spatial parts in momentum space~\cite{Park}:
\begin{eqnarray}
J_{i,{\rm 1b}}^{0}(p^{2}) &=&  g_{V}(p^{2}) - g_{A} \, \frac{{\bf P}
\cdot {\bm \sigma}_{i}}{2 m} + g_{P}(p^{2}) \, \frac{E \, ({\bf p}
\cdot {\bm \sigma}_{i})}{2 m} \,, 
\label{Jt} \\
{\bf J}_{i,{\rm 1b}}(p^{2}) &=& g_{A}(p^{2}) \, {\bm \sigma}_{i}
- g_{P}(p^{2}) \, \frac{{\bf p} \, ({\bf p} \cdot {\bm \sigma}_{i})}{2m}
 + i (g_{M}+g_{V}) \, \frac{{\bm \sigma}_{i}\times {\bf p}}{2 m}
- g_{V} \, \frac{{\bf P}}{2 m} \,,
\label{Js}
\end{eqnarray}
where $E=E_i-E_i', {\bf p}={\bf p}_i-{\bf p}_i'$, and ${\bf P}={\bf
p}_i+{\bf p}_i'$. $g_{V}(p^{2})$, $g_{A}(p^{2})$,
$g_{P}(p^{2})$, and $g_{M}(p^{2})$ are the vector ($V$), axial ($A$), pseudo-scalar
($P$), and magnetic ($M$) couplings. In
chiral EFT, the $p$ dependence is due to loop corrections and
pion propagators, to order~$Q^2$:
$g_{V,A}(p^{2}) = g_{V,A} \, (1-2 \frac{p^2}{\Lambda_{V,A}^2})$, with
$g_V=1$, $\Lambda_V=850 \mev$, 
$\Lambda_A = 2 \sqrt{3}/r_A = 1040 \mev$;
$g_P(p^2) = \frac{2 g_{\pi p n} F_\pi}{m_\pi^2+{\bf p}^2} - 4 \, g_A(p^2)
\frac{m}{\Lambda_A^2}$ and $g_M=\mu_p-\mu_n=3.70$, with pion decay 
constant $F_\pi = 92.4 \mev$, 
$m_{\pi}=138.04 \mev$, and $g_{\pi p n} = 13.05$~\cite{Bernard}.

At leading order $Q^0$, only the momentum-independent $g_A$ and $g_V$
terms contribute. They give rise to low-momentum-transfer Gamow-Teller and Fermi decays. The former has the known form,
$E_1^A|_{\rm LO}\!=\!  \frac{i\,g_A}{\sqrt{6\pi}}\sum_{i=1}^A \sigma_i \tau^{-}_i$. 

Among the $Q^2$ terms, form-factor-type (ff) contributions and the
$g_P$ part of ${\bf J}_{i,{\rm 1b}}$ dominate. The remaining $Q^2$ terms are odd under
parity, so they require either a $P$-wave lepton or another
odd-parity term to connect to $0^+$ states, common for ground states. In such cases, the ${\bf P}$ and
$E$ terms in Eqs.~(\ref{Jt}) and (\ref{Js}) can be neglected, and 
only the term with the large $g_{M}+g_{V} = 4.70$ is
kept.

At order $Q^3$, 2b currents enter in chiral EFT~\cite{Park}. These
include vector spatial, axial temporal, and axial spatial
parts\footnote{Vector temporal parts do not contribute
at this order~\cite{Park}.}. The first two are odd under parity, and
therefore can be neglected in couplings to $0^{+}$ ground states. In addition, within chiral EFT the 
vector spatial current is conserved, satisfying the CVC hypothesis, thus at low-energies their contribution
can be calculated via the Siegert theorem.
As a result, the dominant weak 2b currents include an axial spatial component,
${\bf J}^{\rm axial}_{\rm 2b} = \sum^A_{i<j} {\bf J}_{ij}$,
with~\cite{Park}
\begin{align}
&{\bf J}_{12} = -\frac{g_A}{F^2_{\pi}} \Bigl[ 2 d_1 ({\bm \sigma}_1
\tau_1^- + {\bm \sigma}_2 \tau_2^-) + d_2 \, {\bm \sigma}_{\times} 
\tau^-_{\times} \Bigr] \nonumber \\ 
-&\,\frac{g_A}{2 F^2_{\pi}} \frac{1}{m^2_{\pi}+{\bf k}^2}
\Bigl[ \Bigl(c_4+\frac{1}{4 m} \Bigr) \, {\bf k} \times
({\bm \sigma}_{\times} \times {\bf k}) \, \tau^-_{\times} \nonumber \\ 
+&\,4 c_3 {\bf k} \cdot ({\bm \sigma}_1 \tau_1^- + {\bm \sigma}_2
\tau_2^-) {\bf k} - \frac{i}{2m} {\bf k} \cdot ({\bm \sigma}_1-{\bm
\sigma}_2) {\bf q} \, \tau^-_{\times} \Bigr] ,
\label{2b}
\end{align}
where $\tau_{\times}^-=(\tau_1\times\tau_2)^-$ and the same for ${\bm
\sigma}_{\times}$, ${\bf k}=\frac{1}{2}({\bf p}'_2-{\bf p}_2-{\bf
p}'_1+{\bf p}_1)$ and ${\bf q}=\frac{1}{4}({\bf p}_1+{\bf p}'_1-{\bf
p}_2-{\bf p}'_2)$. Equation~(\ref{2b}) includes contributions
from the one-pion-exchange $c_3, c_4$ parts
and from the short-range couplings $d_1, d_2$, where due to the Pauli
principle only the combination $d_1+2 d_2 = c_D/(g_{A}\Lambda_{\chi})$ 
enters [with $\Lambda_{\chi} = 700 \mev$], see Fig~\ref{currents}.  This result is extremely interesting as it connects the 
low-energy constants (LECs) determining the strength of the meson exchange current (MEC) contact term to $c_D$, which is one of the two unknown low-energy 
constants, $c_D$ and $c_E$, that determine the 3N force up to N$^3$LO, and do not contribute to 2N force. 
This relation shows that one can use a weak observable, e.g., an empirical extraction of 
$\langle E_1^A\rangle|_{emp}$ from a decay rate, as a constraint for the determination of $c_D$ and $c_E$.

\section{Using triton $\beta$-decay to determine the 3N force}
We use the relation between the short range contribution to the current and the 3N force to fully constrain the Lagrangian in systems with up to 3-body, and predict $^4$He properties. The calculation steps are as follows: (i) calculate the $^3$H and $^3$He g.s. wave functions by solving the Schr\"odinger equation for three nucleons interacting via the chiral $NN$ potential at N$^3$LO of Ref.~\cite{N3LO-NN} and the $NNN$ interaction at N$^2$LO~\cite{N2LO-NNN}; (ii) determine for which $c_D$ values along the trajectory the calculated
reduced matrix element of the $E_1^A$ operator (at N$^3$LO) reproduces
$\langle E_1^A\rangle_{emp}$ extracted from $^3$H $\beta$-decay. We find that these two constraints are uncorrelated, producing a stringent 
determination of the LECs: $c_D=-0.2\pm0.1$ and $c_E=-0.205\pm0.015$ \cite{Gazit}.

\begin{table}[t]
\caption{Calculated $^3$H, $^3$He and $^4$He g.s. energies (MeV) and point-proton root-mean-squared radii (fm)}
\begin{tabular}{lcccccc}

\label{predictions}
&  \multicolumn{2}{c}{$^3$H} & \multicolumn{2}{c}{$^3$He} & \multicolumn{2}{c}{$^4$He}\\\cline{2-3}\cline{4-5}\cline{6-7}\\[-2mm]
&  $E_{\rm g.s.}$ & $\langle r^2_p\rangle^{1/2}$ & $E_{\rm g.s.}$ & $\langle r^2_p\rangle^{1/2}$ & $E_{\rm g.s.}$  & $\langle r^2_p\rangle^{1/2}$ \\ [0.7mm]
\hline\\[-3mm]
$NN$& $-$7.852(4) & 1.651(5) & $-$7.124(4) & 1.847(5) & $-$25.39(1)& 1.515(2) \\
$NN\!+\!NNN$&$-$8.473(4) & 1.605(5) & $-$7.727(4) & 1.786(5) & $-$28.50(2) & 1.461(2)\\
Expt. &$-$8.482$\phantom{(5)}$ & 1.60$\phantom{8(5)}$ & $-$7.718\phantom{(1)} & 1.77\phantom{7(1)} & $-$28.296\phantom{()} & 1.467(13)
\end{tabular}
\end{table}
With this calibration of $c_D$ and $c_E$, for this potential,
in principle, any other calculation
is a prediction! In Table~\ref{predictions} a collection of $A\!=\!3$ and 4 data is given, obtained with and without inclusion of the $NNN$ force for $c_D\!=\!-0.2$ ($c_E\!=\!-0.205$). The $NN\!+\!NNN$ predictions for the $^4$He
are in good agreement with measurement.

\begin{figure}[t]
\begin{center}
\rotatebox{0}{\resizebox{10cm}{!}{\includegraphics{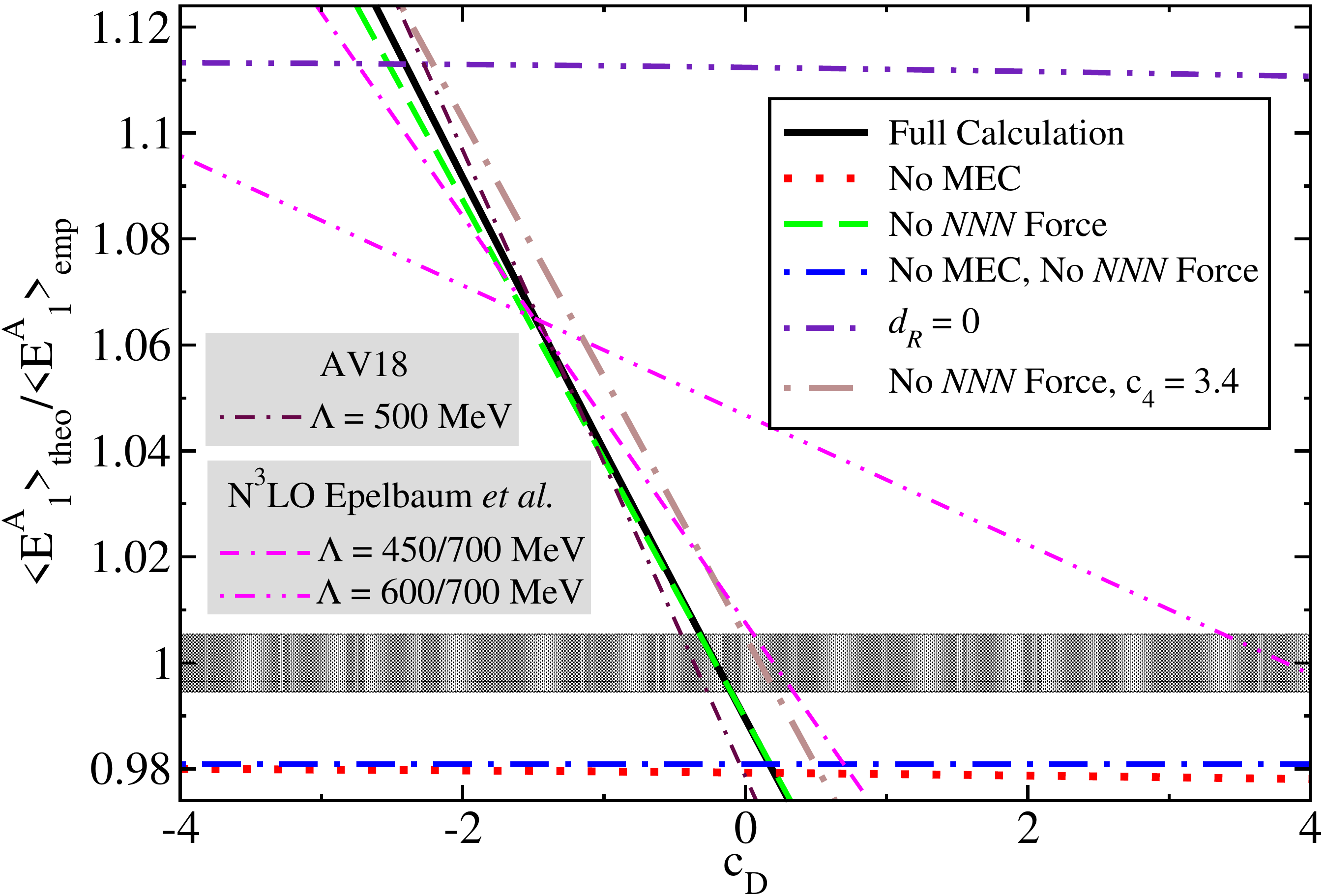}} }
\caption{The ratio $\langle E_1^A\rangle_{th}/\langle E_1^A\rangle_{emp}$
using the N$^3$LO $NN$ potential~\cite{N3LO-NN} with and without the N$^2$LO $NNN$ interaction~\cite{N2LO-NNN},  and the axial current with and without MEC. The shaded area is twice the experimental uncertainty. Also shown: results for the phenomenological AV18 potential (with $\Lambda=$ 500 MeV imposed in the current), and for the N$^3$LO $NN$ potential of Epelbaum et al.~\cite{Epelbaum-NN} (with $\Lambda=450, 600$ MeV, and a 700 MeV spectral-function cutoff in the two-pion exchange), which should be compared with the full calculation (black
solid line). The line denoted by $d_R=0$ includes only the long range contribution to the MEC, neglecting the contact term contribution. Figure taken from \cite{Gazit}.}
\label{Fig:checks}
\end{center}
\end{figure}

To this important conclusion, one can add the results of few additional tests, appearing in Fig.~\ref{Fig:checks}, that are aimed to probe the axial correlations in the nucleus, by analyzing the sensitivity of the triton half life to $NNN$ force and/or MEC:

\begin{itemize}
\item {\it The fundamental importance of the axial two-body currents}.
By suppressing the MEC, in the whole investigated $c_D$-$c_E$ range, the calculations underpredict  $\langle E_1^A\rangle_{emp}$ by about 2\%. Within the MEC, one finds a large cancellation between the long-range one-pion-exchange term (corresponding to the left diagram in Fig.~\ref{currents}), and the contact term (second diagram from the left Fig.~\ref{currents}).   

\item {\it A negligible effect of the $NNN$ force on this observable}.

\item {\it Weak dependence on the specific character of the force}.
\end{itemize}

These features, which might be unique to $s$-shell nuclei, entail that the determination of $c_D$ and $c_E$ obtained in this work is robust.

\section{Extracting the weak structure of the nucleon from $\mu^{-}$ capture on $^3$He \cite{2008PhysLettBGazit}} \label{sec:mu_cap}
In the previous section it was shown that for a weak transition in the 3-nucleon sector, the specific character of the 
force had a minor effect on the reaction rate. We extrapolate this understanding to the muon capture on  $^3\rm{He}$ 
which results in a triton, $\mu^{-} + ^{3}\rm{He} \rightarrow \nu_\mu + ^{3}\rm{H}$. We study this reaction in a
hybrid approach: the calculation uses the chiral EFT based weak current, combined with the phenomenological
nucleon-nucleon potential Argonne $v_{18}$ (AV18) \cite{AV18} augmented by the Urbana IX (UIX) \cite{UIX} 
three nucleon force. This allows a study of the cutoff variation of the observable. 

$\mu$ capture on $^3$He has an accurately measured rate
$\Gamma(\mu^{-} + ^{3}\rm{He} \rightarrow \nu_\mu + ^{3}\rm{H})^{exp}=1496(4) \, \rm{Hz}$,
i.e. a $\pm 0.3\%$ precision \cite{3He_mu_meas}. Combining with the sizable momentum transfer in ordinary
muon capture (OMC), that enhances the effect of the induced-pseudoscalar form factor, allows its study. In addition, one can use OMC to study second class currents \cite{Standard_model_nuclear_tests_RMD}.

The calculation process demands fixing the single contact unknown LEC, which determines the weak axial current.
We do this by reproducing the experimental triton half--life, for various cutoff values for the chiral current. 
The results for the nuclear matrix element of the muon capture process show a $9\%$ effect due to the MEC
contribution, and a small $0.3\%$ effect due to the cutoff dependence. 
It is worthwhile noting that
the relative contribution of the MEC to this process is almost three times bigger than the
MEC contribution to the triton half life. An extremely weak cutoff dependence shows that the
essential physics is captured in the chiral EFT operators.

The final prediction for the capture rate is
\begin{equation} \label{eq:final}
\Gamma= 1499 (2)_\Lambda (3)_{\rm{NM}} (5)_t (6)_{\mathrm{RC}}\, \rm{Hz},
\end{equation}
where the first error is due to the Heavy Baryon chiral perturbation theory (HB$\chi$PT) cutoff, the second is due to
uncertainties in the extrapolation of the form factors to finite momentum
transfer, and in the choice of the specific nuclear model, 
the third error is related
to the uncertainty in the triton half life,
and the last error is due to theoretical uncertainty
in the electroweak radiative corrections calculated for nuclei \cite{MuCap_RC}.
This sums to a total error estimate of about $1\%$, which overlaps with the uncertainty 

The conservative error estimation still allows rather interesting conclusions. First, one notices that
the calculated capture rate agrees with the experimental measurement
$\Gamma(\mu^{-} + ^{3}\rm{He} \rightarrow \nu_\mu + ^{3}\rm{H})^{exp}_{stat}=1496(4)\rm\, \rm{Hz}$. Thus, one
concludes that EFT accurately predicts the capture rate.

However, the most interesting result concerns the weak form factors of the nucleon. In order to constrain the
induced pseudoscalar and second class form factors, we take the following approach. In each case, we set all the other
form factors to their nominal value, and change this form factor in a way which keeps an overlap between the experimental rate
and the theoretically allowed rate. The nominal value of the form factor is set to reproduce the experimental measurement.

The resulting constraint on the induced pseudoscalar form factor is:
\begin{equation}
g_P(q^2=-0.954 m_\mu^2) = 8.13 \pm 0.6 \,,
\end{equation}
in very good agreement with the HB$\chi$PT
prediction of Ref.~\cite{2001PhysRevCKaiser}. 

A second conclusion concerns the contribution of second class currents. While the current constraint on the axial G-
parity breaking term is rather weak, this calculation puts the tightest limit on CVC, constraining it to $m_eF_S/F_V = (0.5 \pm 2) \times 10^{-4}$, which is consistent with CVC. One notes that this is already close to the regime of CVC breaking, predicted by $\chi$PT \cite{2001PhysRevCKaiser}.

A recent calculation in the framework of the previous section, i.e., taking currents and forces from the same chiral
EFT Lagrangian, has reached similar results \cite{Marcucci_2012}.

\section{The in-medium behavior of the axial constant}
Surprisingly, key aspects of $\beta$ decays remain a puzzle. In
particular, when calculations of Gamow-Teller (GT) transitions of the
spin--isospin-lowering operator $g_A {\bm \sigma} \tau^{-}$ are
confronted with experiment, some degree of renormalization, or
``quenching'' $q$, of the axial coupling $g_A^{\rm eff} = q g_A$ is
needed. Compared to the single-nucleon value 
$g_A=1.2695(29)$, the GT term seems to be weaker in nuclei. This was
first conjectured in studies of $\beta$-decay rates, with a typical
$q \approx 0.75$ in many-body calculations~\cite{WildenthalMP,QRPA_EDF}. In view of the significant
effect on weak reaction rates, it is no surprise that this suppression
has been the target of many theoretical works. It is also a
major uncertainty for \znbb\ decay nuclear matrix elements (NMEs),
which probe larger momentum transfers of the order of the pion mass, $p
\sim m_\pi$, where the renormalization could be different. Here we
revisit this puzzle based on chiral EFT currents.

\begin{figure}
\begin{center}
\includegraphics[scale=0.335,clip=]{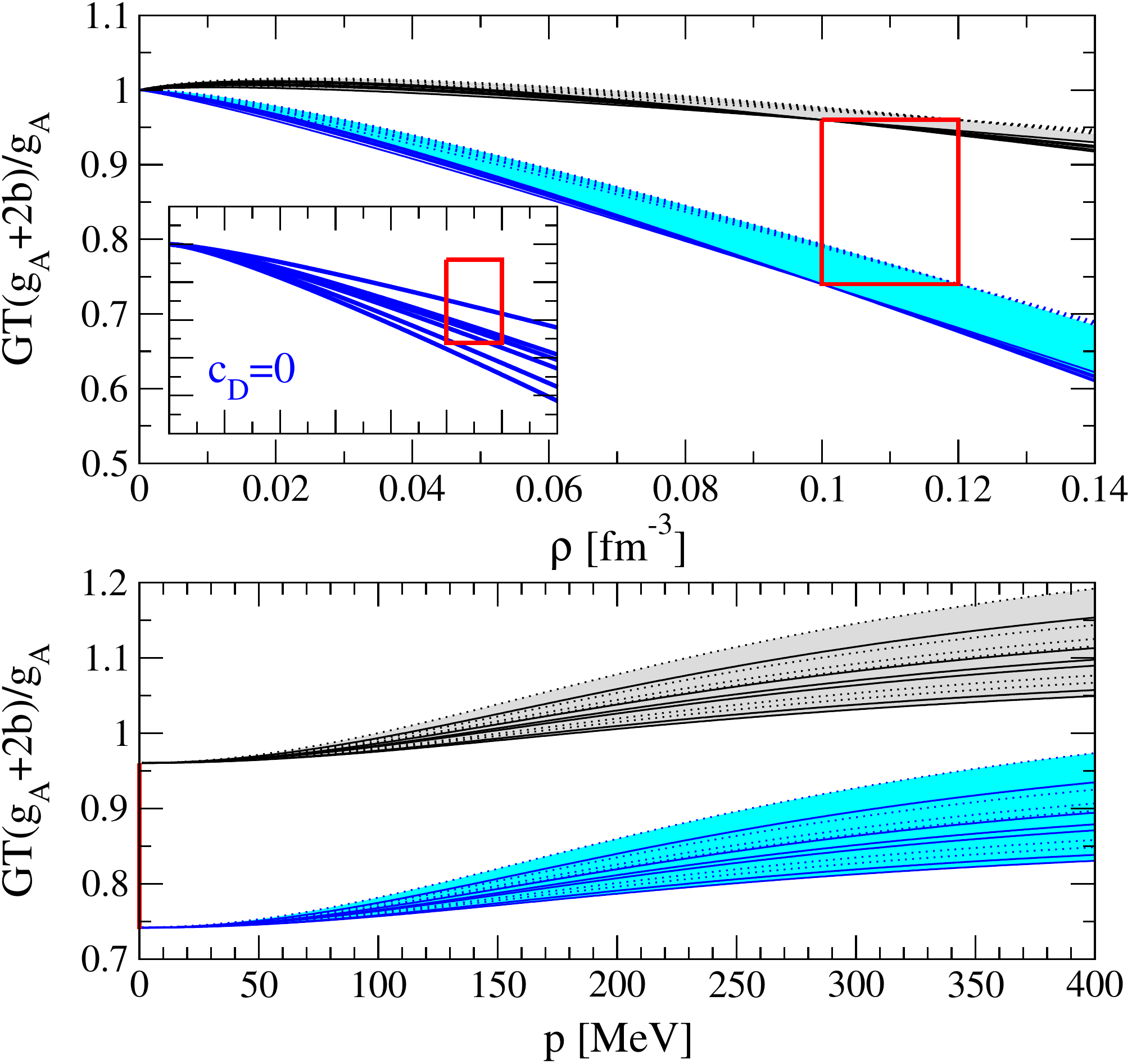}
\end{center}
\caption{ Top panel: $g_A$ plus 2b-current contributions
for $p=0$ GT transitions normalized to $g_A$ as a function of
density $\rho$. The boundaries of the box are given by $q=0.74/0.96$
and $\rho=0.10,...,0.12 \fmiq$. The different curves correspond to possible 
calibrations of $c_D$ (see Ref.~\cite{Javier_0nubb} for details), with shaded regions for the density
range. Intermediate quenching values would lie between these regions.
The inset shows the quenching predicted by the
long-range parts of 2b currents only ($c_D=0$). Bottom panel: Same
as top, but as a function of momentum transfer $p$ for
empirical/smaller quenching $q=0.74/0.96$. Figure taken from Ref.~\cite{Javier_0nubb}.
\label{GT_1b2b}}
\end{figure}

The lightest nucleus that undergoes a $\beta$-decay is the
triton. However, the theory cannot be checked in the triton since its
half-life is used to calibrate the strength of the MEC. The
lightest nucleus that can provide a test to the theory is thus
$^6$He ($\rm{J}^\pi=0^{+}$), an unstable nucleus, which undergoes
a $\beta$ decay with a half-life $\tau_{1/2}=806.7 \pm
1.5\,\rm{msec}$ to the ground state of $^6$Li ($\rm{J}^\pi=1^{+})$
\cite{AjzenbergSelove:1988ec}. As this is a $6$-body problem, we use a
soft nuclear potential, JISP16. This potential leads to an
underbinding of about $80$keV for the $^3$He and $120$keV for the triton, where we 
calibrate the one unknown LEC determining the current. One calculates binding energies of 
$28.70$MeV for $^6$He and $31.46\, \rm{MeV}$ for $^6$Li. These results underestimate the experimental energies by about $0.5
\mev$, though the difference $\Delta E = 2.76$MeV differs by merely $34$keV from the experimental
value. 

Incorporating the $\chi$PT based contributions to the weak-current one can
calculate the full $^6$He-$^6$Li GT matrix-element, as a function of the cutoff imposed in the weak current:
\begin{equation}
|\rm{GT}(^6{\rm He})|_{theo}=2.198 (1)_\Lambda (2)_N (4)_t
(5)_{g_A} = 2.198 \pm 0.007.
\end{equation}
The first error is the cutoff variation dependence, the second is numerical,
the third is due to uncertainties in the triton half-life, and the last is due
to uncertainties in $g_A$. This should be compared to the experimental
matrix-element $|\rm{GT}(^6{\rm He})|_{expt}=2.161\pm 0.005$. Thus, the theory overpredicts
GT by about $1.7\%$, which is reasonable considering the approximation in using a phenomenological, pure
2-body, potential.

For heavier nuclei a microscopic calculation is still out of reach. We thus study the impact of chiral 
2b currents in nuclei at the normal-ordered 1b level by summing the second nucleon over occupied
states in a spin and isospin symmetric reference state or core. Taking a Fermi gas approximation for the core
and neglecting tensor-like terms $({\bf k} \cdot {\bm
\sigma} \, {\bf k} - \frac{1}{3} k^2 {\bm \sigma}) \tau^-$, we obtain
the normal-ordered 1b current \cite{Javier_0nubb}:
\begin{align}
{\bf J}^{\rm eff}_{i,{\rm 2b}} &= 
- g_A {\bm \sigma}_i \tau_i^- \frac{\rho}{F^2_\pi} 
\biggl[ \frac{c_D}{g_A \Lambda_{\chi}} + \frac{2}{3} \, c_3 \,
\frac{{\bf p}^2}{4m^2_{\pi}+{\bf p}^2} + I(\rho,P) \biggl( \frac{1}{3} \, (2c_4-c_3) + \frac{1}{6 m} \biggr)
\biggr] \,,
\label{1beff}
\end{align}
where $\rho = 2 \kf^3/(3\pi^2)$ is the density of the reference state,
$\kf$ the corresponding Fermi momentum, and $I(\rho,P)$ is due to the
summation in the exchange term.
The effective 1b current ${\bf J}^{\rm eff}_{i,{\rm 2b}}$ only
contributes to the GT operator and can be included as a correction to
the $g_A(p^2)$ part of the 1b current, Eq.~(\ref{Js}).  
This demonstrates that chiral 2b currents naturally contribute to the quenching of GT transitions. Compared to light nuclei,
their contributions are amplified because of the larger nucleon
momenta. A quenching of $p \approx 0$ GT transitions is predicted, as well as the momentum transfer dependence, as seen in Fig.~\ref{GT_1b2b}. 

This formalism can be used to evolve the chiral EFT understanding various phenomena, including the prediction
of \znbb\ decay rates. Indeed, such a calculation of the \znbb\
 decay operator based on chiral EFT currents shows that 2b
contributions to the nuclear matrix elements are significant, and range from $-35 \%$ to $10 \%$ \cite{Javier_0nubb}. Another recent application, which will not be covered in this contribution, is to the scattering of a weakly interacting massive
particle (WIMP) candidate on nuclei \cite{Javier_WIMP}.

\section{Summary}
We showed that chiral EFT enables parameter free calculations of electro-weak Gamow Teller transitions, that can be used in various fields: from constraining the nuclear forces, through a deeper understanding of in-medium nuclear effects, and up to predicting weak reactions for the use of experiments studying the limits of the Standard Model.


\section{Acknowledgments}
It is a great pleasure to thank my collaborators in the works presented in this contribution. This work was 
supported by the German Federal Ministry of Education and Research (BMBF) ARCHES -- Award for 
Research Cooperation and High Excellence in Science.

\end{document}